\documentclass[%
pre
,twocolumn%
, notitlepage
,superscriptaddress
,amsmath,amssymb,aps
, nofootinbib
]{revtex4-2}
\usepackage[dvipdfmx]{graphicx}
\usepackage{bm}
\usepackage{dcolumn}
\usepackage{color}
\usepackage{tabularx}
\usepackage{braket}
\usepackage{MnSymbol}
\usepackage{multirow}
\usepackage{xurl}
\usepackage{mathtools}
\usepackage[normalem]{ulem}

\newcommand{\bp}{\overline{\psi}}
\newcommand{\bv}{\overline{v}}
\newcommand{\br}{\boldsymbol{r}}

\newcommand{\rhoi}{\rho}

\renewcommand\sout{\bgroup\color{blue} \ULdepth=-.5ex \ULset}
\graphicspath{{./figs/}}

\allowdisplaybreaks

\begin{document}

\title{Functional renormalization group for classical liquids without recourse to hard-core reference systems: A study of three-dimensional Lennard-Jones liquids}
\author{Takeru Yokota}
\email{takeru.yokota@oit.ac.jp}
\affiliation{Faculty of Engineering, Osaka Institute of Technology, Asahi-ku, Osaka, 535-8585, Japan}
\affiliation{RIKEN Center for Interdisciplinary Theoretical and Mathematical Sciences (iTHEMS), Wako, 351-0198, Japan}
\author{Jun Haruyama}
\email{jun.haruyama@riken.jp}
\affiliation{RIKEN Pioneering Research Institute (PRI), Wako, 351-0198, Japan}
\author{Osamu Sugino}
\email{sugino@issp.u-tokyo.ac.jp}
\affiliation{Institute for Solid State Physics, The University of Tokyo, Kashiwa, Chiba 277-8581, Japan}
\date{\today}

\date{\today}

\begin{abstract}
    In our previous work [Phys.~Rev.~E \textbf{104}, 014124 (2021)], we developed a method for analyzing classical liquids using the functional renormalization group (FRG) without relying on a hard-core reference system. In this paper, we extend this method to three-dimensional liquids. We describe an efficient approach for performing the spatial integrals that appear in the renormalization group equations, which is essential for realizing numerical calculations in three dimensions. As a demonstration, we present its application to the Lennard-Jones liquids. Through calculations of thermodynamic quantities, we find that FRG preserves thermodynamic consistency (TC) better than traditional integral-equation methods such as the hypernetted-chain, Percus-Yevick, and Kovalenko-Hirata closures. Taking the molecular dynamics results as a benchmark, we also show that FRG can achieve an accuracy comparable to that of integral-equation methods that incorporate TC, such as the Rogers-Young closure. We further assess the accuracy of the pair distribution function and examine whether our method remains applicable below the critical temperature. Our results demonstrate that FRG provides a new method for describing classical liquids with accuracy comparable to modern liquid theories.
\end{abstract}
\maketitle

\section{Introduction}
The development of analytical methods for classical liquids has long been a central challenge in statistical mechanics \cite{han13}. In recent years, advances in simulation techniques like molecular dynamics (MD) and Monte Carlo (MC) methods have made it possible to investigate the microscopic state of liquids directly from fundamental principles. However, obtaining highly accurate results still requires substantial computational cost, making it difficult to address slow or large-scale phenomena such as phase transitions. Therefore, the importance of liquid theory remains undiminished, serving as a theoretical framework for systematically understanding the mechanisms behind physical phenomena with a small number of parameters and for rapidly predicting physical properties. The role of liquid theory is not limited to understanding the properties of liquids themselves. For example, it has also evolved into a practical method for efficiently incorporating solvent effects into electronic structure calculations \cite{Nishihara2017}. This method enables the uniform treatment of the distribution of electrons and nuclei (particles) under a single density-based principle.

However, even the theoretical framework for simple liquids is still incomplete. The predictive accuracy of integral equation approaches, such as the hypernetted-chain (HNC) and Percus-Yevick (PY) equations, depends heavily on the precision of the approximation known as the closure relation \cite{Ornstein1914,Percus1958,Morita1960,Baxter1968}.
In addition, such approaches tend to exhibit significant thermodynamic inconsistency—that is, discrepancies arising when the same quantity is computed using different thermodynamic relations \cite{han13}. Attempts to improve predictive power by imposing physical self-consistency (e.g., modified HNC \cite{Rosenfeld1979}, Rogers-Young (RY) closure \cite{Rogers1984}) have achieved some success but have not led to a universal solution. This is also the case in the field of classical density functional theory (cDFT) \cite{Ebner1976,Yang1976,Saam1977,Evans1979}. For repulsive interactions, such as those in hard-sphere systems, fundamental measure theory (FMT) \cite{Rosenfeld1989,Rosenfeld1990_1, Rosenfeld1990_2, Rosenfeld1990_3} has enabled highly accurate descriptions. However, it is still an issue to incorporate attractive interactions essential for real liquids beyond a perturbative treatment \cite{Tang2003}. This issue of incorporating attractive interactions also arises in the rational function approximation (RFA) \cite{Yuste1991, Yuste1996}, another approach that is well suited to hard-core systems. Thus, establishing a general-purpose liquid theory that is both efficient under all conditions and has an accuracy comparable to MD simulations remains an important and ongoing challenge in condensed matter physics.

Here, we focus on the renormalization group as an alternative approach to describing liquids \cite{par85,sal92,par93,whi93,par95,whi95,cai06,par08,par09,cai11,par12,lue15,iso19}. This method has been developed primarily as a framework for treating critical phenomena in liquids. Among such approaches, the well-known hierarchical reference theory (HRT) \cite{par85,par93,par95,par08,par09,par12} takes a system with hard-core repulsion as the reference system and gradually incorporates the effects of the attractive part of the intermolecular interactions, starting from short-range components. This procedure is formulated in terms of differential equations (flow equations), and solving them allows one to analyze the properties of the critical point. Since apparent divergences arise when handling hard-core repulsion within the flow equations, the renormalization group is typically applied only to the attractive part. However, this approach requires prior knowledge of hard-core systems and does not provide a fully consistent analysis based on the flow equations. Moreover, while the renormalization group method should in principle be applicable to liquids beyond just the critical point, studies in this broader context are relatively scarce compared to those focused on critical phenomena.

Against this backdrop, in our previous work \cite{Yokota21} we developed a flow-equation-based method for calculating liquid structures without relying on a hard-core reference system and applicable beyond the critical point. Our approach is grounded in the functional renormalization group (FRG), a formulation based on functionals with a rigorous theoretical foundation. Originally introduced in quantum field theory \cite{weg73,wil74,pol84,wet93} and later applied in the context of quantum systems to density functional theory \cite{pol02,sch04,kem13,Rentrop2015,kem17a,lia18,yok18,yok18b,yok19,yok20,yok21,Yokota2022}, we adapted this framework to construct renormalization group equations for classical systems. From these equations, we derived hierarchical flow equations for multiparticle distribution functions. By reformulating them in terms of cavity distribution functions, we discovered that the apparent divergences arising from hard-core interactions could be eliminated. In that earlier study, we applied the method to an exactly solvable one-dimensional homogeneous system, where it was suggested that FRG offers better accuracy in the high-density regime compared to well-known integral equation methods.

In this paper, we extend our previous work \cite{Yokota21} to apply FRG to three-dimensional homogeneous liquids. As in the earlier study, we introduce a functional flow equation associated with the evolution of the pair interaction. From this, we derive hierarchical flow equations for the cavity distribution functions, which we solve by truncating them using the Kirkwood superposition approximation (KSA) \cite{kir35}. A key step in realizing the application of this method to three-dimensional systems is the efficient evaluation of the spatial integrals appearing in the flow equations. To this end, we introduce a method based on Legendre polynomial expansion \cite{Barker1962} to compute these integrals efficiently. Furthermore, we consider the evolution with respect to the interaction range so as to further reduce the dimensionality of the integrals, which is also a key factor in improving computational efficiency.

In the latter part of this paper, we apply our method to the Lennard-Jones liquid. Here, we test the performance of our approach in the intermediate-density regime. This is done by comparing thermodynamic quantities such as the pressure and the pair distribution function at temperatures above the critical point with the database, such as the modified Benedict-Webb-Rubin (MBWR) equation \cite{Pieprzyk2018} and 
a Span-Wagner-type Helmholtz-energy equation of state (HEOS) \cite{Thol2016}, and integral equation methods. As the density increases, well-known closures, such as the HNC and PY, and the Kovalenko-Hirata (KH) \cite{Kovalenko1999}, exhibit a pronounced breakdown of thermodynamic consistency (TC). In contrast, FRG shows a much smaller violation of TC. Moreover, taking the MD results as a benchmark, we find that the  FRG results of the thermodynamic quantities achieve accuracy comparable to that of the MBWR equation, the HEOS, and the RY closure, which is constructed to enforce TC. We also perform calculations below the critical temperature. In the spinodal region, the bulk compressibility approaches zero and the FRG calculation breaks down, whereas outside this region the FRG remains applicable.

This paper is organized as follows: In Sec.~\ref{sec: formalism}, we present the formulation of our method and introduce an efficient approach for evaluating the three-dimensional spatial integrals appearing in the flow equations. In Sec.~\ref{sec: LJ}, we describe the application of our method to the Lennard-Jones liquid. Section \ref{sec: conclusion} is devoted to the conclusion.

\section{Formalism and technics \label{sec: formalism}}
We introduce our method for analyzing three-dimensional classical liquids. First, we review the approaches presented in our previous work \cite{Yokota21} in Secs.~\ref{sec: flow equation} and \ref{sec: KSA}. We then describe in Sec.~\ref{sec: reduction of spatial integrals} the method developed to extend the analysis to the three-dimensional case.

\subsection{Flow equations for hard-core systems \label{sec: flow equation}}

\subsubsection{Functional flow equation}
We consider a classical particle system interacting via a two-body potential $v_\lambda(\boldsymbol{r} - \boldsymbol{r}')$, where $\boldsymbol{r} - \boldsymbol{r}'$ is the distance between two particles, under an external field $U(\boldsymbol{r})$. Here, a flow parameter $\lambda \in [0, 1]$ is introduced into the two-body interaction. We study the evolution equation with respect to $\lambda$, assuming that $v_\lambda(\boldsymbol{r} - \boldsymbol{r}')$ interpolates between a reference system, $v_0(\boldsymbol{r} - \boldsymbol{r}') = v_{\mathrm{ref}}(\boldsymbol{r} - \boldsymbol{r}')$, and the target system of interest, $v_1(\boldsymbol{r} - \boldsymbol{r}') = v(\boldsymbol{r} - \boldsymbol{r}')$. In the spirit of the renormalization group, the evolution of $v_\lambda(\boldsymbol{r}-\boldsymbol{r}')$ is arranged so that the short-range components of $v(\boldsymbol{r}-\boldsymbol{r}')-v_{\mathrm{ref}}(\boldsymbol{r}-\boldsymbol{r}')$ are incorporated first, followed gradually by the long-range components.

As in cDFT, we introduce the intrinsic free energy density functional \(F_\lambda[\rho]\), which is defined via the grand-canonical distribution and is obtained by subtracting the explicit external-field dependence \(\int d\boldsymbol{r}\rho(\boldsymbol{r}) U(\boldsymbol{r})\) from the Helmholtz free energy density functional \cite{han13}. Here, \(\rho(\br)\) denotes the expectation value of the particle number density in this distribution, hereafter simply referred to as the density. By determining \(F_\lambda[\rho]\), we not only obtain the thermodynamic quantities of the system but also, through its derivatives, the direct correlation functions, thereby gaining access to all information on particle correlations. To define \(F_\lambda[\rho]\), we first introduce the grand partition function \(\Xi_\lambda[\overline{\psi}]\) under inverse temperature \(\beta\) and chemical potential \(\mu\):
\begin{align}
    \Xi_{\lambda}[\overline{\psi}]
    \coloneqq
    &
    \sum_{N=0}^{\infty}
    \int d\br_1\cdots \int d\br_N
    \frac{e^{-\sum_{i<j}^{N}\bv_\lambda(\br_i-\br_j)+\sum_{i=1}^{N}\bp
    (\br_i)}}{\Lambda^{3N} N!}.
\end{align}
Here, $\Lambda$ is the thermal de Broglie wavelength, \(\bv_\lambda(\br_i-\br_j)=\beta v_\lambda(\br_i-\br_j)\) is the dimensionless two-body interaction, and
\(\bp(\br) = \beta (\mu - U(\br))\) is the dimensionless intrinsic chemical potential, obtained by subtracting the explicit contribution of the external field \(\beta U(\br)\) from \(\beta \mu\) \cite{han13}.
The functional $F_\lambda[\rho]$ is the Legendre transform of the thermodynamic potential $\Omega_\lambda[\overline{\psi}] = -\frac{1}{\beta} \ln \Xi_\lambda[\overline{\psi}]$. For notational simplicity, we introduce the dimensionless intrinsic free energy density functional $\overline{F}_\lambda[\rho] = \beta F_\lambda[\rho]$ and the dimensionless thermodynamic potential $\overline{\Omega}_\lambda[\overline{\psi}] = \beta \Omega_\lambda[\overline{\psi}]$. The definition of $\overline{F}_\lambda[\rho]$ is given by
\begin{align}
    \label{eq:def_helm}
    \overline{F}_\lambda[\rho]
    =
    \int d\br
    \rho(\br)
    \overline{\psi}_{\lambda}[\rho](\br)
    +
    \overline{\Omega}_\lambda[\overline{\psi}_{\lambda}[\rho]].
\end{align}
Here, $\overline{\psi}_{\lambda}[\rho](\boldsymbol{r})$ is the function $\overline{\psi}(\boldsymbol{r})$ that maximizes the right-hand side, and therefore satisfies the following condition.
\begin{align}
	\label{eq:F_psimax}
	\rho(\boldsymbol{r})
    =
    -\frac{\delta \overline{\Omega}_{\lambda}[\overline{\psi}_\lambda[\rho]]}{\delta \overline{\psi}(\br)}.
\end{align}
Since the derivative of $-\overline{\Omega}_\lambda[\overline{\psi}]$ gives the expected value of the density, it follows that $\overline{\psi}_{\lambda}[\rho](\boldsymbol{r})$ is the intrinsic chemical potential required to realize the density $\rho(\boldsymbol{r})$.

The FRG equation of $\overline{F}_\lambda[\rho]$ with respect to $\lambda$ is known to be expressible in a closed form as follows \cite{par85,Yokota21}:
\begin{align}
    \label{eq:flow eq}
    &\partial_\lambda
    \overline{F}_\lambda[\rho]
    \notag
    \\
    =&
    \frac{1}{2}
    \int d\boldsymbol{r} \int d\boldsymbol{r}'
    \partial_\lambda \bv_{\lambda}(\br-\br')
    \notag
    \\
    &\times
    \left(
    \rho(\boldsymbol{r})
    \rho(\boldsymbol{r}')
    +
    \left(
    \frac{\delta^2 \overline{F}_\lambda[\rho]}{\delta\rho\delta\rho}
    \right)^{-1}(\boldsymbol{r},\boldsymbol{r}')
    -
    \rho(\boldsymbol{r})
    \delta(\boldsymbol{r}-\boldsymbol{r}')
    \right).
\end{align}
Here, \(\left( \frac{\delta^{2} \overline{F}_{\lambda}[\rho]}{\delta \rho , \delta \rho} \right)^{-1}(\boldsymbol{r}, \boldsymbol{r}')\) is defined in the sense of operator inversion; that is, it is determined by the following equation:
\begin{align}
\int d\boldsymbol{r}''
\frac{\delta^{2} \overline{F}_{\lambda}[\rho]}{\delta \rho(\boldsymbol{r}) \delta \rho(\boldsymbol{r}'')}
\left(
\frac{\delta^{2} \overline{F}_{\lambda}[\rho]}{\delta \rho \delta \rho}
\right)^{-1}(\boldsymbol{r}'', \boldsymbol{r}')
=
\delta(\boldsymbol{r} - \boldsymbol{r}').
\end{align}
Equation \eqref{eq:flow eq} alone determines $F_\lambda[\rho]$ exactly. However, a numerical method for solving such a functional differential equation has not been established, and in practice, some form of approximation is necessary. A mainstream approach involves applying a functional Taylor expansion to the functional differential equation to derive a hierarchy of equations for the expansion coefficients. This hierarchy is then truncated and solved. In fact, such methods have been developed in the context of FRG \cite{dup21}. This approach provides a non-perturbative framework that does not rely on the weakness of the interaction.

To solve the evolution equation \eqref{eq:flow eq}, prior knowledge of the initial reference system is required. The simplest choice for the reference system is the non-interacting case $v_{\mathrm{ref}} = 0$, for which $\Omega_\lambda[\overline{\psi}]$ can be expressed analytically:
\begin{align}
    \overline{F}_0[\rho]
    =
    \int d\boldsymbol{r}
    \rho(\boldsymbol{r})\left(\ln (\Lambda^3\rho(\boldsymbol{r}))-1\right).
\end{align}
The problem arises when $v_\lambda$ interpolates between a non-interacting reference system and a system with hard-core repulsion: the interaction flow $\partial_\lambda v_\lambda$ becomes divergent at short distances, leading to numerical instability. In practice, classical liquid analysis based on flow equations---such as in the HRT---typically treats the hard-core part as the reference system and applies the flow equation only to the long-range attractive part. This kind of approach requires prior knowledge of the hard-core system and prevents computation within a fully self-consistent framework based entirely on the flow equation.

However, this divergence is only apparent, and in approaches based on functional Taylor expansion, the issue can be avoided by reformulating the hierarchy of equations. Specifically, by rewriting the system in terms of a hierarchy for the cavity distribution function, the divergence can be circumvented. This is one of the main results of our previous work \cite{Yokota21}.

\subsubsection{Hierarchical equations for cavity distribution functions}
To define the cavity distribution functions, we first introduce the $n$-particle distribution function:
\begin{align}
	\label{eq:def_g}
	g^{(n)}_\lambda(\br_1,\ldots,\br_n)
	=
	\frac{\rho^{(n)}_\lambda(\br_1,\ldots,\br_n)}{\rho(\br_1)\cdots \rho(\br_n)},
\end{align}
where the $n$-particle density $\rho_\lambda^{(n)}(\boldsymbol{r}_1, \ldots, \boldsymbol{r}_n)$ is given by the following functional derivative expression:
\begin{align}
    \label{eq:def_rhon}
    \rho_\lambda^{(n)}(\boldsymbol{r}_1,\ldots, \boldsymbol{r}_n)
    =&
    \frac{
    e^{\overline{\psi}_{\lambda}[\rho](\boldsymbol{r}_1)}
    \cdots 
    e^{\overline{\psi}_{\lambda}[\rho](\boldsymbol{r}_n)}
    }
    {\Xi_\lambda\left[\overline{\psi}_{\lambda}[\rho]\right]}
    \frac{\delta^n\Xi_\lambda\left[\overline{\psi}_{\lambda}[\rho]\right]}
    {\delta e^{\overline{\psi}(\boldsymbol{r}_1)}\cdots \delta e^{\overline{\psi}(\boldsymbol{r}_n)}}.
\end{align}
The $n$-particle cavity distribution function $y^{(n)}_\lambda(\boldsymbol{r}_1, \ldots, \boldsymbol{r}_n)$ is defined in terms of \(g^{(n)}_\lambda(\br_1,\ldots,\br_n)\) as follows:
\begin{align}
	\label{eq:def_y}
	y^{(n)}_\lambda(\br_1,\ldots,\br_n)
	=
	e^{\sum_{i<j}^{n} \bv_\lambda (\br_i-\br_j)}
	g^{(n)}_\lambda(\br_1,\ldots,\br_n).
\end{align}
Due to this exponential factor, the direct interaction among the $n$ particles is effectively suppressed, and it is known that $y^{(n)}_\lambda(\boldsymbol{r}_1, \ldots, \boldsymbol{r}_n)$ remains a continuous function even in the presence of divergent repulsions or discontinuities in the interaction potential \cite{han13}.

Through Eq.~\eqref{eq:def_helm} and
Eqs.~\eqref{eq:def_g}-\eqref{eq:def_y}, $y^{(n)}_\lambda(\boldsymbol{r}_1, \ldots, \boldsymbol{r}_n)$ is found to be related to the $n$th functional derivative of $\overline{F}_\lambda[\rho]$. Therefore, by differentiating the flow equation \eqref{eq:flow eq} $n$ times, one can derive a corresponding flow equation for $y^{(n)}_\lambda(\boldsymbol{r}_1, \ldots, \boldsymbol{r}_n)$. Here, we consider the case of uniform density \(\rho(\br)=\rho\), which is typically of interest in liquid state theory. In this case, translational symmetry implies that $y^{(n)}_\lambda(\boldsymbol{r}_1, \ldots, \boldsymbol{r}_n)$ depends only on the relative positions $\boldsymbol{r}_1 - \boldsymbol{r}_n, \ldots, \boldsymbol{r}_{n-1} - \boldsymbol{r}_n$. Without loss of generality, we may set $\boldsymbol{r}_n = \boldsymbol{0}$ and write $y^{(n)}_\lambda(\boldsymbol{r}_1, \ldots, \boldsymbol{r}_{n-1})$. In addition, $\overline{\psi}_\lambda(\boldsymbol{r})$ becomes spatially uniform and is denoted by a constant $\overline{\psi}_\lambda$. As derived in Ref.~\cite{Yokota21}, differentiating Eq.~\eqref{eq:flow eq} yields the following flow equation:
\begin{widetext}
\begin{align}
	\label{eq:flow_helm_hom}
	&\partial_\lambda
	\frac{\overline{F}_\lambda}{N}
	=
	-
	\frac{\rho}{2}
	\int d{\br}
	\partial_\lambda
	\left[
	e^{-\bv_\lambda(\br)}
	\right]
	y^{(2)}_\lambda(\br),
    \\
	\label{eq:flow_psi}
	&\partial_\lambda \overline{\psi}_{\lambda}
	=
	-
    \beta K_{T,\lambda}
	\int d{\br}
	\partial_\lambda
	\left[e^{-\bv_\lambda(\br)}\right]
	y^{(2)}_\lambda(\br)
	\left(
	1
	+
	\frac{\rho}{2}
	\int d{\br'}
	\left[
	e^{
	-\bv_\lambda(\br+\br')
	-\bv_\lambda(\br')
	}
	\frac{
	y^{(3)}_\lambda(\br+\br',\br')
	}{y^{(2)}_\lambda(\br)}
	-
	1
	\right]
	\right),
    \\
	\label{eq:flow_y_hom}
	&
	\partial_\lambda \ln y_\lambda^{(n)}(\br_1,\ldots,\br_{n-1})
	\notag
	\\
	&=
	\partial_\lambda \overline{\psi}_{\lambda}
	\left(
	n
	+
	\rhoi
	\int d{\br}
	\left[
	e^{-\sum_{i=1}^{n}\bv_\lambda(\br_i-\br)}
	\frac{y^{(n+1)}_\lambda(\br_1,\ldots,\br_{n-1},\br)}
	{y^{(n)}_\lambda(\br_1,\ldots,\br_{n-1})}
	-
	1
	\right]
	\right)
	+
	\rho
	\int d{\br}
	\partial_\lambda
	\left[e^{
	-\sum_{i=1}^{n}\bv_\lambda(\br_i-\br)
	}\right]
	\frac{y^{(n+1)}_\lambda(\br,\br_1,\ldots,\br_{n-1})}
	{y^{(n)}_\lambda(\br_1,\ldots,\br_{n-1})}
	\notag
	\\
	&
	\quad
	+
	\frac{\rhoi^2}{2}
	\int d\br \int d\br'
	\partial_\lambda\left[e^{-\bv_\lambda(\br-\br')}\right]
	y^{(2)}_\lambda(\br-\br')
	\left[
	e^{
	-\sum_{i=1}^{n}\bv_\lambda(\br_i-\br)
	-\sum_{i=1}^{n}\bv_\lambda(\br_i-\br')
	}
	\frac{
	y^{(n+2)}_\lambda(\br,\br',\br_1,\ldots,\br_{n-1})
	}
	{y^{(2)}_\lambda(\br-\br')
	y^{(n)}_\lambda(\br_1,\ldots,\br_{n-1})}
	-
	1
	\right],
\end{align}
\end{widetext}
where \(N=\rho\int d\boldsymbol{r}\) is the particle number and 
\begin{align}
    \label{eq: modulus}
    K_{T,\lambda}
    =
    \frac{\rho}{\beta}
    \left[
    1
    +
	\rhoi
	\int d{\br}
	\left(
	g^{(2)}_\lambda(\br)
	-
	1
	\right)    
    \right]^{-1}
\end{align}
is the isothermal bulk modulus. Note that $\boldsymbol{r}_n = \boldsymbol{0}$ in Eq.~\eqref{eq:flow_y_hom}. The first-, second-, and third equations correspond to the zeroth, first, and $n$th ($n \geq 2$) functional derivatives of Eq. \eqref{eq:flow eq}, respectively. In all of these equations, $\bv_\lambda(\br)$ appears exclusively inside exponential functions. As a result, even if $\bv_\lambda(\br)$ includes divergent repulsive interactions, such divergences do not manifest in the flow equations themselves. Thanks to this, by setting $\bv_{0}(\br)=0$ we can use the non-interacting case as the initial condition and incorporate the effects of the divergent repulsion by solving the flow equations without encountering apparent divergences. In this case, the initial condition is $y_0^{(n)}(\br_1,\ldots,\br_{n-1})=1$, and the dimensionless excess free energy per particle $\overline{F}^{\mathrm{ex}}/N$ and the dimensionless excess chemical potential $\overline{\mu}^{\mathrm{ex}}$ are obtained as
$\overline{F}^{\mathrm{ex}}/N=\overline{F}_{1}/N-\overline{F}_{0}/N$ and $\overline{\mu}^{\mathrm{ex}}=\overline{\mu}_{1}-\overline{\mu}_{0}$, respectively.

\subsection{Truncation based on the Kirkwood superposition approximation \label{sec: KSA}}
To implement numerical calculations, the hierarchy of equations must be truncated at a finite order. When truncating the hierarchy at order $n$, the $(n+1)$th and $(n+2)$th order distribution functions appearing in Eq.~\eqref{eq:flow_y_hom} must be approximated. In doing so, it is essential to ensure that the following cluster decomposability is preserved:
\begin{align}
    &y_\lambda^{(k)}(\boldsymbol{r}_1,\ldots,\boldsymbol{r}_{k-1})
    \notag
    \\
    &\approx
    y_\lambda^{(l)}(\boldsymbol{r}_1,\ldots,\boldsymbol{r}_{l-1})
    \notag
    \\
    &\times
    y_\lambda^{(k-l)}(\boldsymbol{r}_{l+1}-\boldsymbol{r}_{l},\ldots,\boldsymbol{r}_{k-1}-\boldsymbol{r}_{l})
    \notag
    \\
    &\text{when }
    |\boldsymbol{r}_{1,\ldots, l-1}|\ll |\boldsymbol{r}_{l,\ldots, k-1}|.
\end{align}
In other words, when a group of $l$ particles is sufficiently far from the remaining $k - l$ particles, the $k$-body distribution function factorizes into the product of the $l$-body and $(k - l)$-body distribution functions. If this property does not hold, some of the spatial integrals appearing in Eqs. \eqref{eq:flow_psi} and \eqref{eq:flow_y_hom} will diverge, as the integrands will not decay to zero at infinity.

The KSA is one such approximation that satisfies the cluster decomposability condition. In this approximation, many-body distribution functions are expressed as products of lower-order distribution functions, which is justified in regimes where many-body correlations are weak, such as in the low-density limit. For example, by neglecting three-body and higher-order correlations, $y^{(k)}_\lambda$ can be approximated as a product of two-body distribution functions as follows:
\begin{align}
    \label{eq: KSA def}
    y^{(k)}_\lambda(\boldsymbol{r}_1,\ldots, \boldsymbol{r}_{k-1})
    \approx
    \prod_{i=1}^{k-1}\prod_{j=i+1}^{k}
    y^{(2)}_\lambda(\boldsymbol{r}_j-\boldsymbol{r}_i)+O(\rho).
\end{align}
Here, $\boldsymbol{r}_k = \boldsymbol{0}$. 

Since Eq.~\eqref{eq: KSA def} is guaranteed only in the low-density limit, one might expect that applying the KSA to the flow equation would significantly degrade its accuracy at high densities, just as applying it to the Yvon-Born-Green (YBG) equation is known to cause large errors in the virial coefficients \cite{Grouba2004}. However, we do not necessarily expect this to be the case. In our approach, the integration over $\lambda$ incorporates higher-order correlations that become important at high densities, allowing partial inclusion of such correlations even when the right-hand side of the flow equation is approximated by the KSA. Furthermore, unlike the YBG equation, the \(n\)th-order equation depends not only on the \((n+1)\)th-order correlation function but also on the \((n+2)\)th-order one, suggesting that higher-order correlations can be more effectively captured. In addition, even when approximations are applied to the cavity distribution functions, the exponential factors in the integrands of the flow equations ensure that one of the key contributions—the excluded-volume effect—is properly taken into account.

In this work, we consider the flow equations up to second order and apply the KSA to $y^{(3)}_\lambda$ and $y^{(4)}_\lambda$. As a result, Eq.~\eqref{eq:flow_psi} and Eq.~\eqref{eq:flow_y_hom} for $n = 2$ are approximated as follows:
\begin{widetext}
\begin{align}
	\label{eq:flow_psi_KSA}
	\partial_\lambda \overline{\psi}_{\lambda}
	\approx&
	-
    \beta K_{T,\lambda}
	\int d{\br}
	\partial_\lambda f_\lambda(\br)
	y^{(2)}_\lambda(\br)
	\left(
	1
	+
	\frac{\rho}{2}
	\int d{\br'}
	\left(
    g^{(2)}_\lambda(\br+\br')
    g^{(2)}_\lambda(\br')
	-
	1
	\right)
	\right),
    \\
	\label{eq:flow_y_KSA}
	\partial_\lambda \ln y_\lambda^{(2)}(\br_1)
	\approx&
	\partial_\lambda \overline{\psi}_{\lambda}
	\left(
	2
	+
	\rhoi
	\int d{\br}
	\left(
    g^{(2)}_\lambda(\br_1-\br)
    g^{(2)}_\lambda(\br)
	-
	1
	\right)
	\right)
	+
	2\rho
	\int d{\br}
	\partial_\lambda f_\lambda(\br)
    y^{(2)}_\lambda(\br)
    g^{(2)}_\lambda(\br_1-\br)
    \notag
	\\
	&
	\quad
	+
	\frac{\rhoi^2}{2}
	\int d\br \int d\br'
	\partial_\lambda f_\lambda(\br-\br')
	y^{(2)}_\lambda(\br-\br')
	\left(
	g^{(2)}_\lambda(\br)
	g^{(2)}_\lambda(\br')
    g^{(2)}_\lambda(\br-\br_1)
    g^{(2)}_\lambda(\br'-\br_1)
	-
	1
	\right).
\end{align}
\end{widetext}
Here, we have introduced $f_\lambda(\boldsymbol{r}) = e^{-\overline{v}_\lambda(\boldsymbol{r})} - 1$. In the derivation, note that $\boldsymbol{r}_2 = \boldsymbol{0}$.

\subsection{Reduction of spatial integrals\label{sec: reduction of spatial integrals}}

\subsubsection{Legendre expansion}
Eqs.~\eqref{eq:flow_psi_KSA} and \eqref{eq:flow_y_KSA} contain double spatial integrals of the following form:
\begin{align}
    I(r_1)=&\int d\boldsymbol{r}_2\int d\boldsymbol{r}_3
    F_{1}(\boldsymbol{r}_1,\boldsymbol{r}_2)
    F_{2}(\boldsymbol{r}_2,\boldsymbol{r}_3)
    F_{3}(\boldsymbol{r}_3,\boldsymbol{r}_1).
\end{align}
Here, $F_{i}(\boldsymbol{r}, \boldsymbol{r}')$ (\(i=1,2,3\)) is assumed to depend only on $r = |\boldsymbol{r}|$, $r' = |\boldsymbol{r}'|$, and $\cos\theta = \frac{\boldsymbol{r} \cdot \boldsymbol{r}'}{rr'}$; that is, $F_{i}(\boldsymbol{r}, \boldsymbol{r}') = F_{i}(r, r', \cos\theta)$. Reducing the computational cost of this integral is crucial for the feasibility of numerical implementation. To achieve this, we adopt a method based on Legendre expansion \cite{Barker1962}. Using this approach, the spatial integral above can be simplified to the following form of a double integral:
\begin{align}
    \label{eq: I result}
    I(r_1)=&\sum_{l=0}^\infty \frac{16\pi^2}{(2l+1)^2}
    \int_0^\infty dr_2\int_0^\infty dr_3 r_2^2 r_3^2
    \notag
    \\
    &\times
    \tilde{F}_{1,l}(r_1,r_2)\tilde{F}_{2,l}(r_2,r_3)\tilde{F}_{3,l}(r_3,r_1).
\end{align}
Here, $\tilde{F}_{i,l}(r, r')$ denotes the coefficients of the Legendre polynomial expansion of $F_i(r, r', \cos\theta)$, and is defined using the Legendre polynomial $P_l(t)$ as follows:
\begin{align}
    \tilde{F}_{i,l}(r,r')
    =&
    \frac{2l+1}{2}\int_{-1}^{1}dt F_{i}(r,r',t)P_l(t).
\end{align}
When applying Eq.~\eqref{eq: I result} to Eq.~\eqref{eq:flow_y_KSA}, an infinite sum appears. For practical computation, however, a cutoff is introduced to truncate the series.

\subsubsection{Choice of the evolution of interaction}
Depending on the choice of $\bv_\lambda(r)$, the dimensionality of the integrals can be further reduced. Here, we consider a flow in terms of the interaction range, defined as follows:
\begin{align}
    \label{eq: v flow}
    \bv_\lambda(r)=\bv(r)\theta(\lambda r_\mathrm{cut}-r).
\end{align}
Here, we assume that the interaction is short-ranged, and that there exists a cutoff $r_{\mathrm{cut}}$ beyond which the interaction becomes negligibly small. Under Eq.~\eqref{eq: v flow}, $\partial_\lambda f_\lambda(r)$ appearing in the flow equations can be expressed in terms of a delta function as follows:
\begin{align}
    \partial_\lambda f_\lambda(r)
    =
    r_\mathrm{cut}f(\lambda r_\mathrm{cut})\delta(\lambda r_\mathrm{cut}-r).
\end{align}
This appears within the integrand of the flow equations and, as a result, reduces the dimensionality of the integrals. The flow equations after applying Eqs.~\eqref{eq: I result} and \eqref{eq: v flow} take the following form:
\begin{widetext}
\begin{align}
    \label{eq: flow F final}
    \partial_\lambda\frac{\overline{F}_\lambda}{N}
    =&
    -2\pi\lambda^2\rho r_{\mathrm{cut}}^3 f(\lambda r_{\mathrm{cut}}) y^{(2)}_\lambda(\lambda r_{\mathrm{cut}}),
    \\
    \label{eq: flow psi final}
    \partial_\lambda \bp_{\lambda}
    \approx&
    2\frac{\partial_\lambda\overline{F}_\lambda}{N}
    \left(
    1
    +
    2\pi\beta K_{T,\lambda}
    \int_0^{\infty} dr\, r^{2}\tilde{h}_{\lambda,0}(\lambda r_{\mathrm{cut}},r)h^{(2)}_\lambda(r)
    \right),
    \\
    \label{eq: flow y final}
    \partial_\lambda \ln y_\lambda^{(2)}(r_1)
    \approx&
    -4\frac{\partial_\lambda\overline{F}_\lambda}{N}
    \left(
    \tilde{h}_0(\lambda r_{\mathrm{cut}},r_1) 
    +
    2\pi\rho
    \int_{0}^\infty dr' 
    r^{\prime 2}
    \tilde{h}_{\lambda,0}(\lambda r_{\mathrm{cut}},r')\tilde{h}_{\lambda,0}(r',r_1)\left(1+2h_{\lambda}(r')\right)
    \right.
    \notag
    \\
    &
    \qquad\qquad
    \left.
    -
    2\pi\beta K_{T,\lambda}
    \int_0^{\infty} dr\, r^{2}\tilde{h}_{\lambda,0}(\lambda r_{\mathrm{cut}},r)h^{(2)}_\lambda(r)
    \times 
    2\pi\rho\int_0^{\infty} dr'\, r^{\prime 2} h_\lambda(r')\tilde{h}_0(r',r_1)
    \right)
    \notag
    \\
    &
    +
    8\pi^2 \rho^2\sum_{l=0}^{\infty}\frac{1}{(2l+1)^2}
    \int_{0}^\infty dr \int_{0}^\infty dr' r^2 r^{\prime 2}
    h^{(2)}_\lambda(r)\tilde{h}_{\lambda,l}(r,r_1)
    \tilde{Y}_{\lambda,l}(r,r')
    h^{(2)}_\lambda(r')\tilde{h}_{\lambda,l}(r',r_1),
\end{align}
\end{widetext}
where $h_\lambda(r) = g_\lambda^{(2)}(r) - 1$ is the total correlation function, $\tilde{h}_{\lambda, l}(r, r')$ is the coefficient of the Legendre polynomial expansion of $h_\lambda(|\boldsymbol{r} - \boldsymbol{r}'|)$, and $\tilde{Y}_{\lambda, l}(r, r')$ is the coefficient of the Legendre polynomial expansion of $Y_\lambda(\boldsymbol{r} - \boldsymbol{r}') = \partial_\lambda f_\lambda(\boldsymbol{r} - \boldsymbol{r}')\, y_\lambda(\boldsymbol{r} - \boldsymbol{r}')$, given by
\begin{align}
    \tilde{Y}_{\lambda, l}(r,r')
    =&
    \frac{2l+1}{2rr'}
    \lambda r_{\mathrm{cut}}^2
    f(\lambda r_{\mathrm{cut}})y^{(2)}_\lambda(\lambda r_{\mathrm{cut}})
    \notag
    \\
    &\times
	P_l\left(\frac{r^2+r^{\prime 2}-\lambda^2 r_{\mathrm{cut}}^2}{2rr'}\right)
    \notag
    \\
    &\times
    \theta(r+r'-\lambda r_{\mathrm{cut}})
    \theta(\lambda r_{\mathrm{cut}}-|r-r'|).
\end{align}
Equations \eqref{eq: flow F final}-\eqref{eq: flow y final} represent the final forms used in our numerical calculations.

We now discuss the impact on accuracy when using Eq.~\eqref{eq: v flow}. Without introducing approximations, the results would be independent of the specific choice of the \(\lambda\)-evolution of \(\bv_\lambda(r)\); however, the introduction of the KSA introduces such dependence. In the case of incorporating hard-core repulsion, the use of Eq.~\eqref{eq: v flow} ensures accuracy in the initial stage of the flow. This is because the dimensionless parameter reduces to the packing fraction \(\eta = \pi\rho (\lambda r_{\mathrm{cut}})^3 / 6\), where \(\lambda r_{\mathrm{cut}}\) is smaller than the particle diameter for the hard-core system, allowing the \(\lambda\)-evolution to be reinterpreted as a density evolution. Consequently, for small \(\lambda\), the KSA is guaranteed to hold, and the flow equation becomes accurate.

\section{Analysis of Lennard-Jones fluid \label{sec: LJ}}
\begin{figure*}
    \centering
    \includegraphics[width=1.0\linewidth]{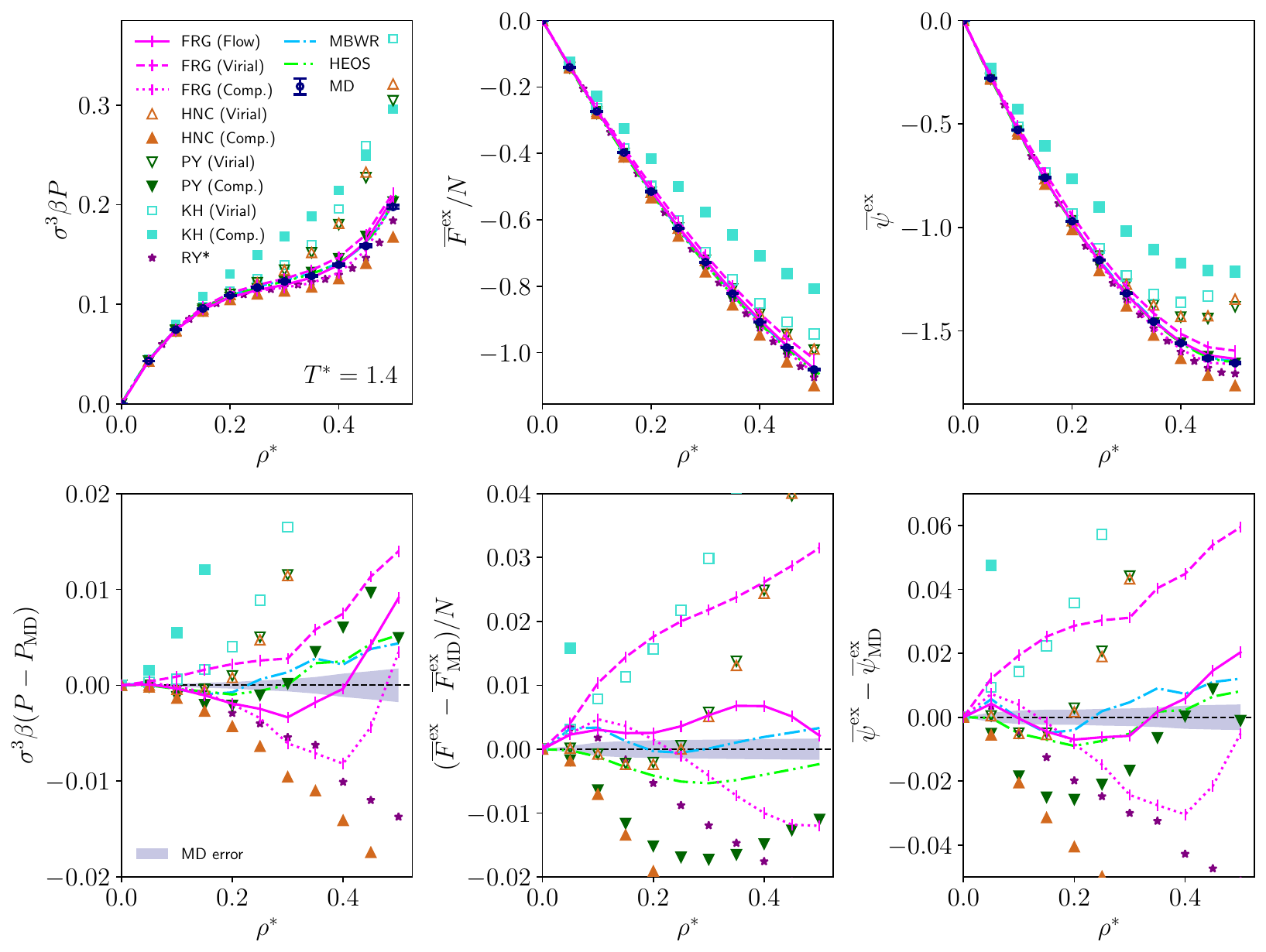}
    \caption{(Upper panels) Density dependence of the pressure \(P\), excess free energy \(\overline{F}^{\mathrm{ex}}/N\), and excess chemical potential \(\overline{\psi}^{\mathrm{ex}}\) at \(T^{*} = 1.4\) for FRG, the integral equation methods, the MBWR equation, the HEOS and MD. (Lower panels) Deviations of the results obtained by each method from the MD values of the pressure \(P_{\mathrm{MD}}\), excess free energy \(\overline{F}^{\mathrm{ex}}_{\mathrm{MD}}/N\), and excess chemical potential \(\overline{\psi}^{\mathrm{ex}}_{\mathrm{MD}}\).
    For both the FRG and the integral equation methods (HNC, PY, and KH), results obtained via the virial route [Eq.~\eqref{eq: p virial}] and compressibility route [Eq.~\eqref{eq: p comp}] are presented and labeled as (Virial) and (Comp.), respectively. Note that for the RY\(^*\) closure these two routes coincide, and therefore no separate labels are assigned. In addition, for the FRG, the results obtained from the flow-equation route [Eqs.~\eqref{eq: flow F final} and \eqref{eq: flow psi final}] are also shown and labeled as (Flow). The MD results are obtained via the virial route. The shaded regions in the lower panels correspond to the error bars of the MD results in the upper panels. For the error estimation, see Appendix \ref{app: MD integral}.
    }
    \label{fig: thermo}
\end{figure*}
\begin{figure}
    \centering
    \includegraphics[width=\linewidth]{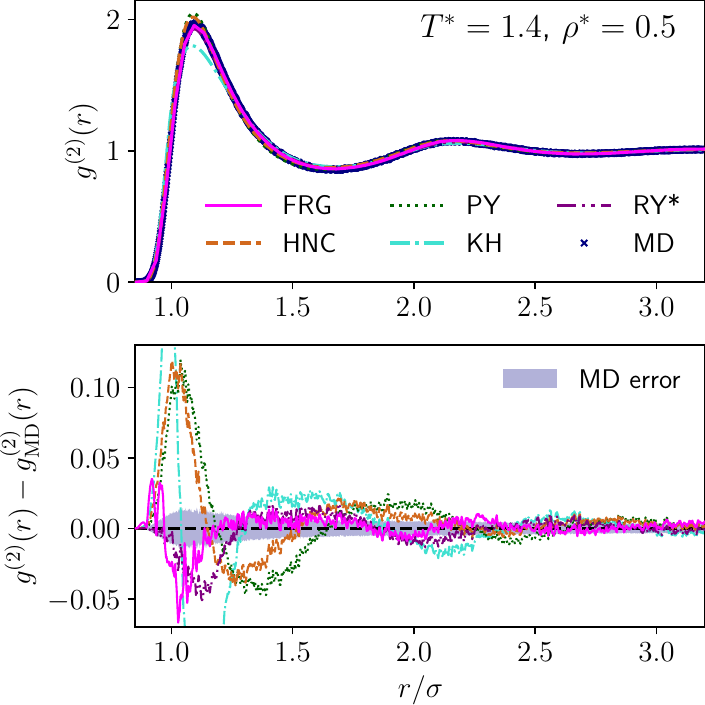}
    \caption{Pair distribution function results from FRG, integral equation methods (HNC, PY, KH, RY\(^*\)), and MD at $T^*=1.4$, $\rho^*=0.5$. The lower panel shows the differences between the MD result $g^{(2)}_{\mathrm{MD}}(r)$ and those of the other methods. The shaded region in the lower panel indicates the error of the MD results. For the error estimation, see Appendix \ref{app: MD integral}.}
    \label{fig: dist0.5}
\end{figure}
\begin{figure}
    \centering
    \includegraphics[width=\linewidth]{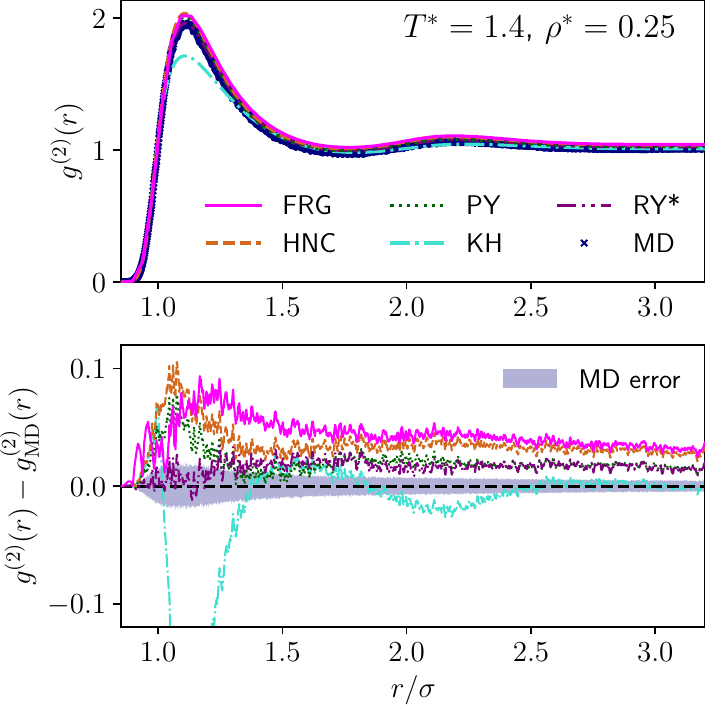}
    \caption{Same as Fig.~\ref{fig: dist0.5}, but for $\rho^*=0.25$.}
    \label{fig: dist0.25}
\end{figure}
Here, we apply our method to the Lennard-Jones (LJ) liquid, one of the standard models used to evaluate the performance of methods for classical liquids. The interaction in the LJ liquid is expressed in terms of the depth $\epsilon$ and particle diameter $\sigma$ as follows:
\begin{align}
    v(r)
    =
    4\epsilon\left(
    \left(\frac{\sigma}{r}\right)^{12}
    -
    \left(\frac{\sigma}{r}\right)^{6}
    \right).
\end{align}
Using these two parameters, the state of the system is specified by the dimensionless density $\rho^*=\rho\sigma^3$ and the dimensionless temperature $T^*=1/(\beta\epsilon)$. One of the notable properties of the LJ model is that it exhibits a critical point associated with the liquid-gas phase transition. The critical point is located at $(\rho^*, T^*)=(\rho_c^*, T_c^*)\approx(0.3, 1.3)$ \cite{han13}, and for $T^*<T_c^*$, the system undergoes a first-order phase transition as $\rho^*$ increases.

In particular, we address the following two points:
\begin{enumerate}
    \item For $T^*>T_c^*$, where calculations can be stably performed over a wide range of $\rho^*$, we compare our results with those of MD, the integral equation methods, the modified Benedict-Webb-Rubin (MBWR) equation \cite{Pieprzyk2018}, and a Span-Wagner-type Helmholtz-energy equation of state (HEOS) \cite{Thol2016} in order to assess the accuracy and the thermodynamic consistency of our approach.
    \item For \( T < T_c^* \), we investigate how the FRG works in the presence of a phase transition.
\end{enumerate}
Details of the numerical setup for FRG are presented in Appendix \ref{app: numerical}, while the MD simulations, integral equation calculations, the MBWR equation, and the HEOS are described in Appendix \ref{app: MD integral}.

\subsection{Results for \(T^*>T_c^*\)}
Here, we present the results for thermodynamic quantities and the pair distribution function \(g^{(2)}(r)=g_{1}^{(2)}(r)\) at $T^*=1.4 > T_c^*$. We also compare these results with those obtained from MD, the MBWR equation, the HEOS, and the integral equation methods. For the integral equation approaches, we employ the following four closures: the hypernetted-chain (HNC) \cite{Morita1960,Baxter1968}, the Percus-Yevick (PY) closure \cite{Percus1958}, the Kovalenko-Hirata (KH) closure \cite{Kovalenko1999}, and the Rogers-Young closure \cite{Rogers1984} with the renormalized indirect correlation function \cite{Weeks1971} (RY\(^*\)). The RY\(^*\) closure is determined so as to ensure thermodynamic consistency (see Appendix \ref{app: MD integral}) and is believed to describe the properties of the Lennard-Jones liquid more accurately than the HNC or PY closures \cite{Pihlajamaa2024}.

In addition to accuracy, we examine thermodynamic consistency---that is, the agreement between results obtained from different methods of calculating thermodynamic quantities. To this end, we compute the excess free energy \(\overline{F}^{\mathrm{ex}}/N = \overline{F}_{1}/N - \overline{F}_{0}/N\), the excess chemical potential \(\overline{\psi}^{\mathrm{ex}} = \overline{\psi}_{1} - \overline{\psi}_{0}\), and the pressure \(P\) using the following three different methods and compare the discrepancies among them:
\begin{enumerate}
    \item (Flow-equation route) By solving Eqs.~\eqref{eq: flow F final} and \eqref{eq: flow psi final}, \(\overline{F}^{\mathrm{ex}}/N\) and \(\overline{\psi}^{\mathrm{ex}}\) are obtained, and \(P\) is then calculated as
    \begin{align}
        \label{eq: pressure}
        \beta P
        =&
        \rho\left(1+\overline{\psi}^{\mathrm{ex}}-\frac{\overline{F}^{\mathrm{ex}}}{N}\right).
    \end{align}
    \item (Virial route) Using \(g^{(2)}(r)\) obtained from FRG, integral equation methods, or MD as input, \(P\) is calculated at several densities \(\rho\) from the pressure equation
    \begin{align}
        \label{eq: p virial}
        \beta P
        =&
        \rho - \frac{2\pi \rho^2}{3} \int_0^{\infty} dr\, r^3
        \frac{d\overline{v}(r)}{dr} g^{(2)}(r).
    \end{align}
    Using the resulting \(P\), \(\overline{F}^{\mathrm{ex}}/N\) is then calculated from 
    \begin{align}
        \label{eq: F integral}
        \frac{\overline{F}^{\mathrm{ex}}}{N}
        =&
        \int_0^{\rho} d\rho'\frac{1}{\rho'}\left(\frac{\left. \beta P \right|_{\rho=\rho'}}{\rho'} - 1\right),
    \end{align}
    where the density integral is evaluated numerically. The excess chemical potential \(\overline{\psi}^{\mathrm{ex}}\) is then calculated using Eq.~\eqref{eq: pressure}.
    \item (Compressibility route) Using \(g^{(2)}(r)\) obtained from FRG or integral equation methods as input, the bulk modulus (inverse compressibility) \(K_T = K_{T,1}\) is calculated at several densities \(\rho\) using Eq.~\eqref{eq: modulus}. The pressure \(P\) is then obtained from
    \begin{align}
        \label{eq: p comp}
        \beta P
        =&
        \int_{0}^{\rho} d\rho'\,
        \frac{\left. \beta K_{T} \right|_{\rho=\rho'}}{\rho'},
    \end{align}
    where the density integral is evaluated numerically. As in the virial route, \(\overline{F}^{\mathrm{ex}}/N\) and \(\overline{\psi}^{\mathrm{ex}}\) are calculated from the resulting \(P\) using Eqs.~\eqref{eq: pressure} and \eqref{eq: F integral}. 
\end{enumerate}
For the MD simulations, we employ only the virial route, since in the compressibility route the integrand in Eq.~\eqref{eq: modulus} can exhibit a long tail, resulting in significant finite-size effects. As in the case of the integral equation approaches, when the MBWR equation of state is used, \(\overline{F}^{\mathrm{ex}}/N\) and \(\overline{\psi}^{\mathrm{ex}}\) are evaluated using Eqs.~\eqref{eq: pressure} and \eqref{eq: F integral}. Note that for the RY\(^*\) closure, no difference arises between the virial and compressibility routes, since the closure is determined so as to make them coincide.

Figure~\ref{fig: thermo} shows the results for \(P\), \(\overline{F}^{\mathrm{ex}}/N\), and \(\overline{\psi}^{\mathrm{ex}}\). Since this region is close to the critical point, the slope of \(P\) with respect to density becomes gentle around \(\rho^* = 0.3\). In terms of accuracy, all methods yield results close to those of MD in the low-density region. 
However, as the density increases, the results obtained from the virial and compressibility routes for the HNC, PY, and KH closures exhibit large discrepancies, indicating a severe breakdown of thermodynamic consistency. In contrast, for FRG, the discrepancy between these routes is much more strongly suppressed. Regarding accuracy, although the RY\(^*\) closure yields better results for \(P\) around \(\rho^{*} = 0.5\), FRG achieves accuracy comparable to that of the MBWR equation and the HEOS for \(P\) via any route. For \(\overline{F}^{\mathrm{ex}}/N\) and \(\overline{\psi}^{\mathrm{ex}}\), FRG obtained via the flow-equation route shows accuracy comparable to that of the MBWR equation and the RY\(^*\) closure.

Figure~\ref{fig: dist0.5} shows the pair distribution function \(g^{(2)}(r)\) at \(\rho^{*} = 0.5\), together with the deviations of each method from the MD results, \(g^{(2)}(r) - g^{(2)}_{\mathrm{MD}}(r)\). Notably, when MD is taken as the reference, FRG reproduces the peak height around \(r/\sigma = 1\) more accurately than the HNC, PY, and KH closures, and with accuracy comparable to that of the RY\(^*\) closure. On the other hand, we also observe that whether FRG yields superior accuracy depends on the density. As seen in Fig.~\ref{fig: dist0.25}, which shows the results for \(\rho^{*} = 0.25\), the RY\(^*\) closure provides higher accuracy than FRG, and no clear superiority of FRG over the HNC or PY closures is observed. Compared with the KH closure, FRG reproduces the peak around \(r/\sigma = 1\) more accurately, whereas the behavior at larger \(r\) is better captured by the KH closure.

\subsection{Results for \(T^*<T_c^*\)}

\begin{figure}
    \centering
    \includegraphics[width=\linewidth]{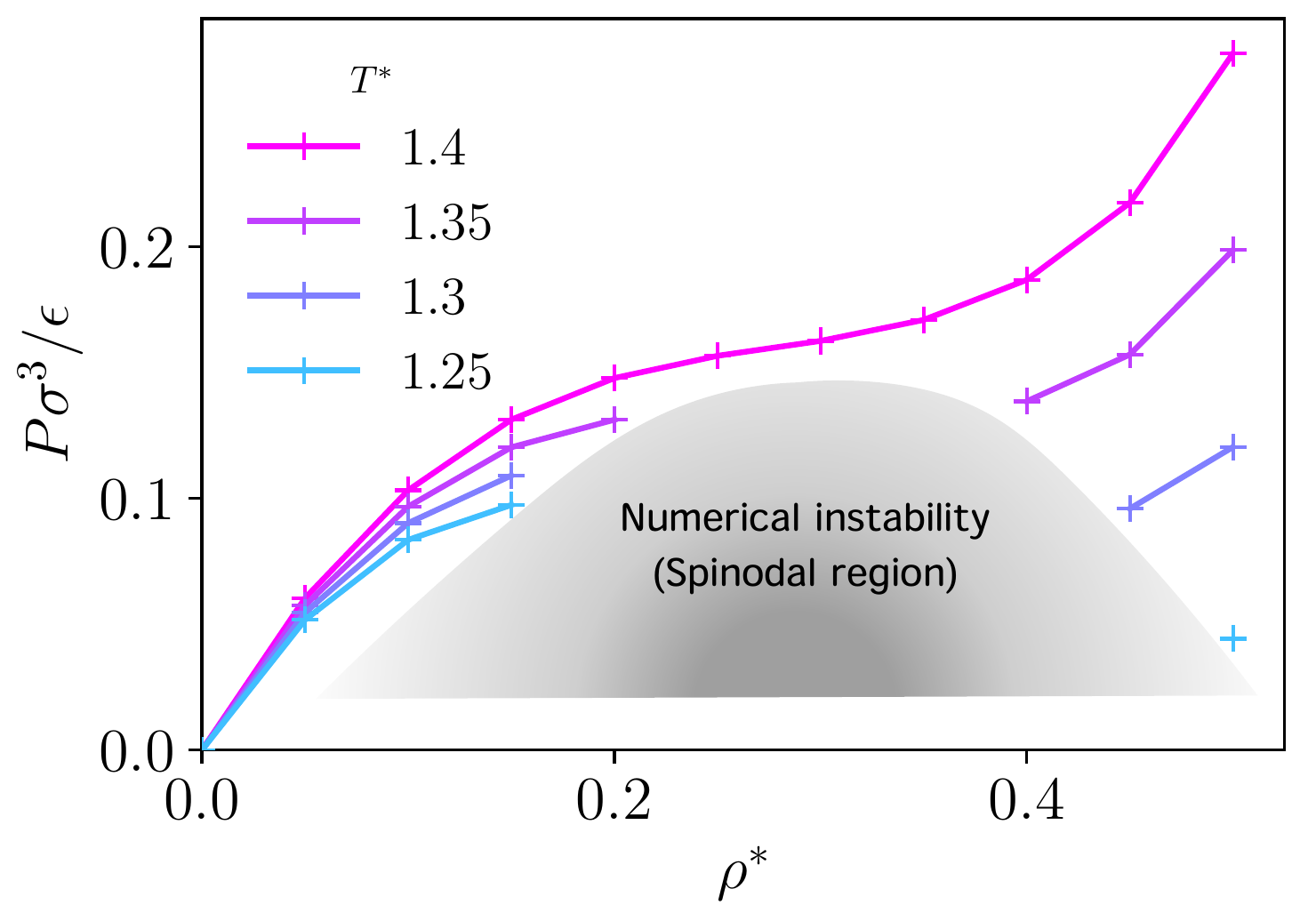}
    \caption{Density dependence of the pressure obtained from FRG. The gray region corresponds to the parameter range in which the bulk modulus approaches zero during the flow and the calculation becomes unstable, suggesting that it is a spinodal region.}
    \label{fig: pressure}
\end{figure}
We examine how the FRG works in the presence of a phase transition below the critical temperature \(T^* < T_c^*\). Figure~\ref{fig: pressure} shows the density dependence of the pressure for various values of \(T^*\). As the system approaches the critical point, the pressure curve becomes flatter. This behavior indicates that the bulk modulus approaches zero, suggesting the emergence of a region where the integral in Eq.~\eqref{eq: modulus} diverges, leading to numerical instability. In fact, we find that such breakdowns start to appear around \(T^* \sim 1.35\). In the figure, only the data points for which the calculations remain stable are plotted, and the region where the calculations break down is indicated by the gray shading.

For \(T^* < 1.35\), the presence of a gas-liquid first-order phase transition implies the coexistence of low-density (gas) and high-density (liquid) states with the same temperature and pressure. Indeed, when we compare data points near the boundary of the numerically unstable region at \(T = 1.25\) and \(T = 1.3\), we find that the pressure becomes lower at higher densities. This behavior, similar to that obtained in mean-field analyses \cite{han13}, indicates the presence of metastable homogeneous states. The density dependence of the pressure observed in this region is a characteristic feature of approximate calculations; as the accuracy of the approximation improves and the convexity of the free energy is restored, the pressure curve should become flat. The region where the bulk modulus reaches zero during the flow, leading to numerical breakdown, expands as the temperature decreases. This behavior suggests that uniform-density states become unstable, which can be interpreted as the emergence of a spinodal region.

\begin{figure}
    \centering
    \includegraphics[width=\linewidth]{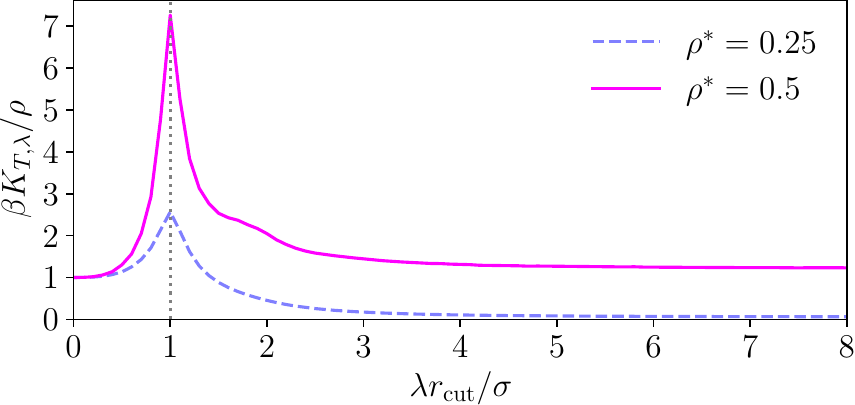}
    \caption{$\lambda$-dependence of the isothermal buk modulus (Eq.~\eqref{eq: modulus}) at $T^*=1.4$ and $\rho^*=0.25, 0.5$.}
    \label{fig: modulus T1.4}
\end{figure}

\subsection{Calculation for higher-density regime}
So far, we have presented calculations for $\rho^*\leq 0.5$. At higher densities, however, we encounter numerical breakdown. This can be understood from the evolution of the bulk modulus $K_{T,\lambda}$. Figure \ref{fig: modulus T1.4} shows the results for $K_{T,\lambda}$ at $T^*=1.4$ and $\rho^*=0.25, 0.5$. As can be seen from the choice of $\overline{v}_\lambda(r)$ in Eq.~\eqref{eq: v flow}, the short-range repulsive part of the LJ potential is incorporated when $\lambda r_{\mathrm{cut}}/\sigma < 1$. Consequently, $K_{T,\lambda}$ increases in this regime, and the increase becomes more pronounced at higher densities. At still larger densities, $K_{T,\lambda}$ grows even more rapidly, which, as is evident from Eqs.~\eqref{eq: flow psi final} and \eqref{eq: flow y final}, leads to a divergence of the flow.

This issue is expected to be avoidable by modifying the choice of flow. In the flow defined by Eq.~\eqref{eq: v flow}, only the repulsive part is incorporated in the early stage of the evolution. By altering this so that the attractive part is also included to some extent, the rapid increase of $K_{T,\lambda}$ can be prevented. Calculations for the high-density regime based on this idea will be presented in a separate paper.

\section{Conclusion \label{sec: conclusion}}
In this paper, we presented a method for analyzing three-dimensional liquids using the functional renormalization group without relying on prior knowledge of a hard-core reference system. As a concrete demonstration, we calculated thermodynamic quantities and the pair distribution function of the Lennard-Jones liquid. We found that our method preserves thermodynamic consistency (TC) much more effectively than integral equation methods that are not based on TC, such as the hypernetted-chain, Percus-Yevick, and Kovalenko-Hirata closures, and exhibits accuracy comparable to that of the modified Benedict-Webb-Rubin equation, the Span-Wagner-type Helmholtz-energy equation of state, and the Rogers-Young closure, which ensures TC. We also performed calculations below the critical temperature. We found that, in the spinodal region, the bulk compressibility approaches zero and the FRG calculation breaks down.

In this study, we focused on the analysis near the critical point and did not perform calculations in the higher-density regime closer to realistic conditions. As discussed above, we expect that such calculations will become feasible in the future by modifying the choice of interaction evolution. Once a method applicable over a broader density range is established, we aim to extend our approach to realistic solvent systems such as water, with the prospect of contributing to practical chemical calculations.

Further improving computational accuracy within the present framework is another direction for future work. For example, instead of using the KSA, it remains to be investigated whether third- and fourth-order correlations can be determined directly from the flow equations. Another possibility is to incorporate known correction terms to the KSA \cite{Abe1959} to enhance accuracy.

\begin{acknowledgments}
    T.~Y.~was supported by JSPS KAKENHI Grant Nos.~25H00177 and 25K23364.    
\end{acknowledgments}

\appendix
\section{Details of numerical calculation \label{app: numerical}}
\begin{figure}
    \centering
    \includegraphics[width=1.0\linewidth]{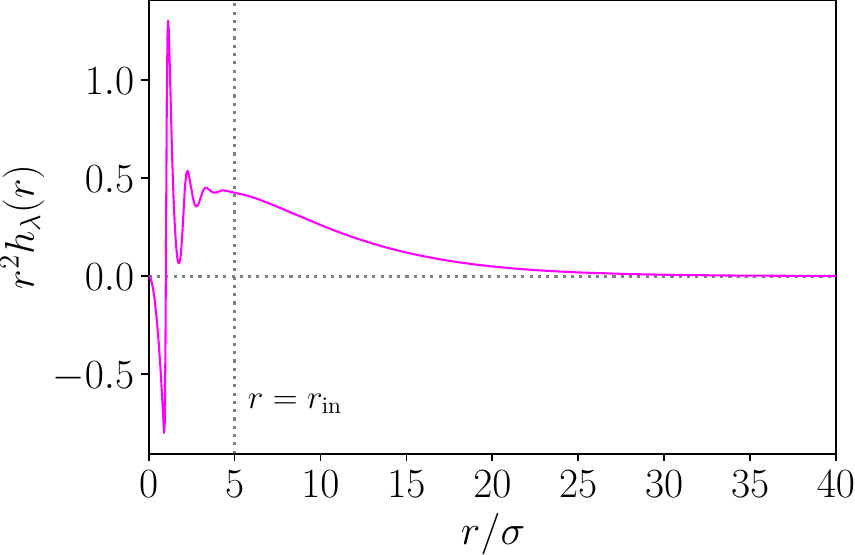}
    \caption{The \(r\)-dependence of the integrand \(r^{2} h_{\lambda}(r)\) in Eq.~\eqref{eq: modulus radial}. The parameters are set to \((\rho^{*}, T^{*}) = (0.25, 1.4)\) and \(\lambda r_{\mathrm{cut}} / \sigma = 8\). The vertical dotted line indicates \(r_{\mathrm{in}} = 5\sigma\).}
    \label{fig: r2h}
\end{figure}
\begin{figure}
    \centering
    \includegraphics[width=1.0\linewidth]{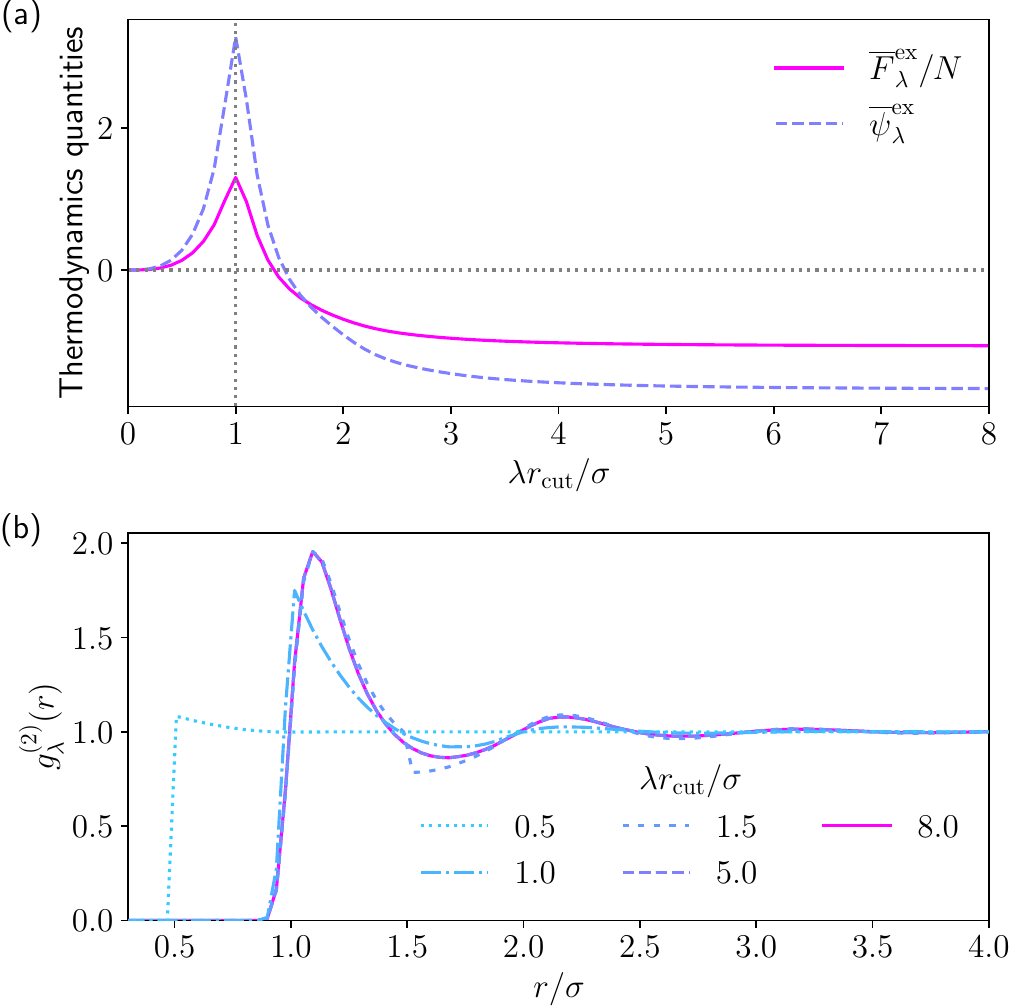}
    \caption{$\lambda$-dependence of $\overline{F}_\lambda^{\mathrm{ex}}/N$, $\overline{\psi}_\lambda^{\mathrm{ex}}$, and $g_\lambda^{(2)}(r)$ in FRG, obtained for $(\rho^*,T^*)=(0.5,1.4)$.}
    \label{fig: flow}
\end{figure}
Here we describe the details of the numerical solution of Eqs.~\eqref{eq: flow F final}-\eqref{eq: flow y final}. 

To treat $y_\lambda^{(2)}(r)$, we discretize $r$ as follows. In the relatively small range $0 \leq r < r_{\mathrm{in}} = 5\sigma$, where $y^{(2)}(r)$ varies significantly, we take an equally spaced grid of $N_{\mathrm{in}}=128$ points, with grid points $r(n)$ ($n=0,\ldots,N_{\mathrm{in}}-1$) defined by
\begin{align}
    \label{eq: grid in}
    r(n)
    =
    \frac{r_{\mathrm{in}}}{N_{\mathrm{in}}} n.
\end{align}
To handle larger values of $r$ without introducing too many grid points, we employ a non-uniform grid in the range $r_{\mathrm{in}} \leq r < r_{\mathrm{out}} = 100\sigma$, where the grid spacing increases at larger $r$ values, reflecting the weaker variation of $y^{(2)}_\lambda(r)$ in that region. Specifically, we take $N_{\mathrm{out}}=64$ grid points $r(n)$ ($n = N_{\mathrm{in}}, \ldots, N_{\mathrm{in}}+N_{\mathrm{out}}-1$) defined by
\begin{align}
    r(n)
    =&
    \frac{1}{N_{\mathrm{out}}^2}
    \left(r_{\mathrm{out}}-\frac{r_{\mathrm{in}}N_{\mathrm{out}}}{N_{\mathrm{in}}}-r_{\mathrm{in}}\right)
    (n-N_{\mathrm{in}})^2
    \notag
    \\
    &+
    \frac{r_{\mathrm{in}}}{N_{\mathrm{in}}}(n-N_{\mathrm{in}})+r_{\mathrm{in}}.
\end{align}
This construction ensures continuity with Eq.~\eqref{eq: grid in} at $n = N_{\mathrm{in}}$, satisfying both the positional continuity $r(N_{\mathrm{in}})=r_{\mathrm{in}}$ and the grid-spacing continuity $r'(N_{\mathrm{in}})=r_{\mathrm{in}}/N_{\mathrm{in}}$, while also fulfilling $r(N_{\mathrm{out}})=r_{\mathrm{out}}$ and covering the region $[r_{\mathrm{in}}, r_{\mathrm{out}})$. Handling such a wide range of $r$ is important near the critical point for evaluating $K_{T,\lambda}$, which appears in the flow equations and is expressed from Eq.~\eqref{eq: modulus} as
\begin{align}
    \label{eq: modulus radial}
    K_{T,\lambda}
    =
    \frac{\rho}{\beta}
    \left(
    1
    +
	4\pi \rhoi
	\int_0^{\infty} dr\,
    r^2 h_\lambda(r)\right)^{-1}.
\end{align}
In fact, as shown in Fig.~\ref{fig: r2h}, the integrand
$r^2 h_\lambda(r) = r^2[e^{-\overline{v}_\lambda(r)}y^{(2)}_\lambda(r)-1]$
can exhibit a long tail extending beyond $r=r_{\mathrm{in}}$.

When evaluating the integrals in Eqs.~\eqref{eq: flow psi final}, \eqref{eq: flow y final}, and \eqref{eq: modulus radial}, which are computed numerically using Gaussian quadrature, the values of $y_\lambda^{(2)}(r)$ at points other than the grid are required. For $0 \leq r \leq r(N_{\mathrm{in}}+N_{\mathrm{out}}-1)$, we use values obtained by linear interpolation between grid points, while for larger $r$ we set $y_\lambda^{(2)}(r)=1$.

For the cutoff $r_{\mathrm{cut}}$ appearing in Eq.~\eqref{eq: v flow}, we set $r_{\mathrm{cut}}=5\sigma$. Figure \ref{fig: flow} shows the evolution of $\overline{F}_\lambda^{\mathrm{ex}}/N$, $\overline{\psi}_\lambda^{\mathrm{ex}}$, and $g_\lambda^{(2)}(r)$ obtained by solving the flow equations. With $r_{\mathrm{cut}}=8\sigma$, these quantities are seen to converge at $\lambda=1$. 

The remaining numerical details are as follows. For the infinite sum over $l$ in Eq.~\eqref{eq: flow y final}, we introduce a cutoff $L_{\max}=10$ and confirm the convergence of the results. To solve the $\lambda$-evolution of Eqs.~\eqref{eq: flow F final}-\eqref{eq: flow y final}, we employ an eighth-order Runge-Kutta method. The implementations of Gaussian quadrature, linear interpolation of $y^{(2)}_\lambda(r(n))$, and the Runge-Kutta method are carried out using the Gnu Scientific Library \cite{Galassi2009}. Furthermore, we accelerate the computations by parallelizing the evaluation of Eq.~\eqref{eq: flow y final} at different grid points $r_1=r(n)$ with OpenMP.

\section{Molecular dynamics, integral equation methods, and the MBWR equation \label{app: MD integral}}
Here, we summarize the setup of the MD calculations and the integral equation methods used in this paper.

\subsection{Molecular dynamics}
We describe the procedure of our MD analysis. First, we present the details of the atomic configuration calculations in the MD simulations, and then we describe the methods used for finite-size corrections and error estimation.

We perform atomic configuration calculations for argon atoms in a cubic box with periodic boundary conditions using the PIMD package \cite{Shiga2025}. The simulations are carried out with \( N = 256, 512, 1024, 2048,\) and \( 4096 \) atoms, and the box length is determined according to the target density \( \rho \). The time evolution is performed up to \( 0.25\,\mathrm{ns} \) with a time step of \( 0.5\,\mathrm{fs} \), corresponding to a total of \( 5 \times 10^5 \) steps. Atomic configurations are stored every \( 1000 \) steps, i.e., every \( 0.5\,\mathrm{ps} \), resulting in a total of \( 500 \) saved configurations. A massive Nos\'e-Hoover thermostat chain is employed to control the temperature.

For each \( N \) and \( \rho \), we estimate the statistical errors of \(g^{(2)}(r)\) and \(P\) using the Flyvbjerg-Petersen (FP) method \cite{Flyvbjerg1989}. This is done by dividing the \(450\) saved configurations, excluding the first \(50\), into \(45\) blocks of \(10\) configurations each. In this procedure, \( g^{(2)}(r) \) is obtained by constructing a histogram of interparticle distances from the configurations, using bins of width \( 0.0016\sigma \) for \( r < 3.35\sigma \). The pressure is calculated using the virial route [Eq.~\eqref{eq: p virial}].

\begin{figure}
    \centering
    \includegraphics[width=1.0\linewidth]{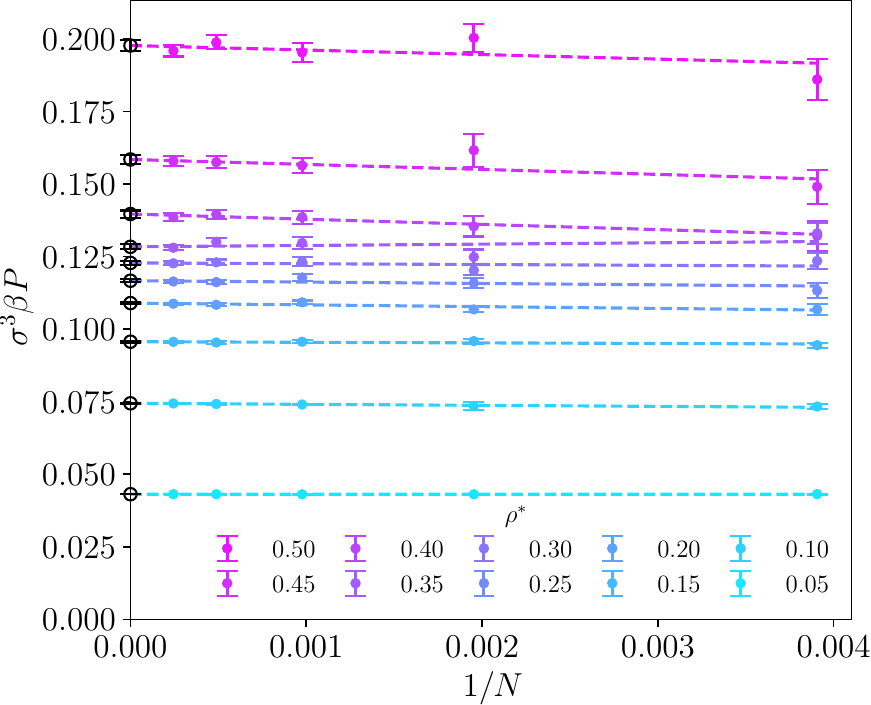}
    \caption{Dependence of \( \sigma^3 \beta P \) on \( 1/N \) at various densities and \(T^*=1.4\). The results for \( N = 256, 512, 1024, 2048, \) and \( 4096 \) are shown as points with error bars, and the dashed lines represent the corresponding weighted linear fits. The extrapolated values for the infinite system are indicated by black circles with error bars at \( 1/N \to 0 \).}
    \label{fig: extra}
\end{figure}
The finite-size effects are corrected by extrapolating the obtained \( g^{(2)}(r) \) and \( P \) for each \( N \) and \( \rho \) to the limit \(N \to \infty\). In all cases, we assume that the finite-size effects scale as \( 1/N \), and we therefore treat each quantity as a function of \( 1/N \) and perform a linear fit. The value in the infinite-system limit is then estimated from the intercept at \( 1/N \to 0 \). In this procedure, we use a weighted linear fit with errors from the FP method to assess the uncertainty of the extrapolated values. As an example, the extrapolation result for \( P \) is shown in Fig.~\ref{fig: extra}. The black circles with error bars at \( 1/N \to 0 \) represent the final results for \( P \) at each density. The same analysis is performed for \( g^{(2)}(r) \) at each value of \( r \), yielding the results with error estimates.

Based on the extrapolated pressure \(\tilde{P}\) and its error \(\Delta P\), we estimate the values and uncertainties of \(\overline{F}^{\mathrm{ex}}\), obtained from the density integral in Eq.~\eqref{eq: F integral}, and \(\overline{\psi}^{\mathrm{ex}}\), obtained via Eq.~\eqref{eq: pressure}. For this purpose, we employ a Monte Carlo error propagation in which \( P \) is sampled as a Gaussian random variable with mean \( \tilde{P} \) and standard deviation \( \Delta P \). In a single Monte Carlo trial, \( P \) is sampled for each \( \rho \), and \( \overline{F}^{\mathrm{ex}} \) and \( \overline{\psi}^{\mathrm{ex}} \) are calculated from the sampled values. We repeat this procedure \(10000\) times to obtain statistical ensembles of \( \overline{F}^{\mathrm{ex}} \) and \( \overline{\psi}^{\mathrm{ex}} \), and estimate their values and errors from the corresponding means and standard deviations. The values and error bars of the thermodynamic quantities shown in Fig.~\ref{fig: thermo} are based on this analysis.

For \( g^{(2)}(r) \), the \(r\)-dependence of the results obtained using the FP method tends to be noisy. To suppress this noise, we apply a Savitzky-Golay filter for smoothing. The window length for the smoothing is set to 11 bins, and a third-order polynomial is used. Figures~\ref{fig: dist0.5} and \ref{fig: dist0.25} are based on the smoothed results. The error bars shown in these figures correspond to those obtained from the FP method.


\subsection{Integral equation method}
The integral equation method is based on the following Ornstein-Zernike equation \cite{Ornstein1914}:
\begin{align}
    \tilde{c}(|\boldsymbol{k}|)
    =
    \frac{\tilde{h}(|\boldsymbol{k}|)}{1+\rho \tilde{h}(|\boldsymbol{k}|)},
\end{align}
where $\tilde{c}(|\boldsymbol{k}|)$ and $\tilde{h}(|\boldsymbol{k}|)$ are the three-dimensional Fourier transforms of the direct correlation function $c(|\boldsymbol{r}|)$ and the total correlation function $h(|\boldsymbol{r}|)=g^{(2)}(|\boldsymbol{r}|)-1$, respectively.

To solve the Ornstein-Zernike equation, integral equation methods introduce a relation known as a closure, which is written as
\begin{align}
g(r) = e^{-\overline{v}(r) + \gamma(r) + b(r)} .
\end{align}
Here, \(\gamma(r) = h(r) - c(r)\) is the indirect correlation function, and \(b(r)\) is called the bridge function. The exact form of \(b(r)\) is generally unknown, and the accuracy of the method is determined by how it is approximated. In this paper, we employ the following four closures.
\begin{enumerate}
    \item Hypernetted chain (HNC) \cite{Morita1960,Baxter1968}: \(b(r)\) is approximated as
    \begin{align*}
        b(r)\approx 0.
    \end{align*}
    \item Percus-Yevick equation (PY) \cite{Percus1958}: \(b(r)\) is approximated as
    \begin{align*}
        b(r)\approx \ln(1+\gamma(r))-\gamma(r).
    \end{align*}
    \item Kovalenko-Hirata closure (KH) \cite{Kovalenko1999}: \(b(r)\) is approximated as
    \begin{align*}
        b(r)\approx \left[\ln(1+d(r))-d(r)\right]\theta\left(d(r)\right),
    \end{align*}
    where \(d(r)=-\overline{v}(r)+\gamma(r)\).
    \item Rogers-Young closure \cite{Rogers1984} with the renormalized indirect correlation function (RY\(^*\)): \(b(r)\) is approximated as
    \begin{align*}
        b(r) \approx \ln\left( 1 + \frac{e^{f(r)\gamma^{*}(r)} - 1}{f(r)} \right) - \gamma^{*}(r),
    \end{align*}
    where \(f(r) = 1 - e^{-\alpha r}\) with a free parameter \(\alpha\), and the renormalized indirect correlation function is defined as \(\gamma^{*}(r) = \gamma(r) - \overline{v}_{\mathrm{LR}}(r)\), with \(\overline{v}_{\mathrm{LR}}(r)\) being the long-range part of the interaction. To ensure thermodynamic consistency, the parameter \(\alpha\) is chosen numerically so as to satisfy
    \begin{align*}
        \frac{\beta K_{T}}{\rho}
        =
        \frac{\partial}{\partial \rho}
        \left[
        \rho - \frac{2\pi \rho^{2}}{3}
        \int_{0}^{\infty} dr r^{3}
        \frac{d\overline{v}(r)}{dr} g^{(2)}(r)\right],
    \end{align*}
    which is obtained by requiring that \(\partial (\beta P)/\partial \rho\) derived from the virial route [Eq.~\eqref{eq: p virial}] and from the compressibility route [Eq.~\eqref{eq: p comp}] be equal. Following the splitting proposed by Weeks et al.~\cite{Weeks1971}, as in Ref.~\cite{Pihlajamaa2024}, we define the long-range part of the interaction as \(\overline{v}_{\mathrm{LR}}(r) = \overline{v}(r) - \overline{v}(r^{*})\), where \(r^{*} = 2^{1/6}\sigma\) is the position at which the Lennard-Jones potential attains its minimum.
\end{enumerate}
In Ref.~\cite{Pihlajamaa2024}, it is suggested that the RY\(^*\) closure describes the properties of the Lennard-Jones liquid more accurately than the HNC or PY closures.

In the numerical calculations for HNC, PY, KH, we used a grid of 6851 points for $r$ over the range $0 \leq r \leq 50\sigma$. The solutions were obtained using an iterative method. Calculations using the RY\(^*\) closure were performed with the open-source code OrnsteinZernike.jl \cite{Pihlajamaa2024}.

\subsection{The MBWR equation}
The MBWR equation is an analytic form of the equation of state for the Lennard-Jones liquid with 33 adjustable parameters and has the following form:
\begin{align}
    &\frac{\sigma^3}{\epsilon} P(\rho,T)
    \notag
    \\
    =&
    \rho^* T^*
    + \rho^{*2}\left(x_{1}T^* + x_{2}\sqrt{T^*} + x_{3}+ \frac{x_{4}}{T^*}+ \frac{x_{5}}{T^{*2}}\right)
    \notag
    \\
    &+
    \rho^{*3}\left(x_{6}T^* + x_{7}+ \frac{x_{8}}{T^*}+ \frac{x_{9}}{T^{*2}}\right)
    \notag
    \\
    &
    + \rho^{*4}\left(x_{10}T^* + x_{11}+ \frac{x_{12}}{T^*}\right)
    + \rho^{*5} x_{13}
    + \rho^{*6}\left(\frac{x_{14}}{T^*}+ \frac{x_{15}}{T^{*2}}\right)
    \notag
    \\
    &
    +\rho^{*7}\frac{x_{16}}{T^*} 
    +\rho^{*8}\left(\frac{x_{17}}{T^*}+ \frac{x_{18}}{T^{*2}}\right)
    +\rho^{*9}\frac{x_{19}}{T^{*2}}
    \notag
    \\
    &
    +
    \Bigg[\rho^{*3}\left(\frac{x_{20}}{T^{*2}}+\frac{x_{21}}{T^{*3}}\right)
    + \rho^{*5}\left(\frac{x_{22}}{T^{*2}}+ \frac{x_{23}}{T^{*4}}
    \right)
    \nonumber
    \\
    &
    +
    \rho^{*7}\left(\frac{x_{24}}{T^{*2}}+ \frac{x_{25}}{T^{*3}}\right)+ \rho^{*9}\left(\frac{x_{26}}{T^{*2}}+ \frac{x_{27}}{T^{*4}}\right)
    \notag
    \\
    &
    +
    \rho^{*11}\left(\frac{x_{28}}{T^{*2}}+ \frac{x_{29}}{T^{*3}}\right)+ \rho^{*13}\left(\frac{x_{30}}{T^{*2}}+ \frac{x_{31}}{T^{*3}}+ \frac{x_{32}}{T^{*4}}\right)\Bigg] e^{-\gamma \rho^{*2}}.
\end{align}
Here, \(x_{1}, \ldots, x_{32}\) and \(\gamma\) are adjustable parameters. We use the parameters proposed in Ref.~\cite{Pieprzyk2018}, where \(\gamma = 3\) is fixed and \(x_{1}, \ldots, x_{32}\) are determined by fitting to MD results, as listed in Table~I of that reference.

\subsection{The HEOS}
The Span-Wagner-type Helmholtz-energy equation of state for Lennard-Jones liquids proposed in Ref.~\cite{Thol2016} is constructed on the basis of molecular simulation data for the Helmholtz free energy and its derivatives with respect to temperature and density, and is given by
\begin{align}
    \frac{\overline{F}^{\mathrm{ex}}}{N}
    =&
    \sum_{i=1}^{6} n_i \delta^{d_i} \tau^{t_i}
    +
    \sum_{i=7}^{12} n_i \delta^{d_i} \tau^{t_i} e^{-\delta^{l_i}}
    \notag
    \\
    &+
    \sum_{i=13}^{23} n_i \delta^{d_i} \tau^{t_i}
    e^{-\eta_i (\delta - \epsilon_i)^2 - \beta_i (\tau - \gamma_i)^2},
\end{align}
where it is defined as a function of the temperature \(T\) and density \(\rho\), with \(T_{\mathrm{c}} = 1.32\) and \(\rho_{\mathrm{c}} = 0.31\), in terms of the reduced variables \(\delta = \rho / \rho_{\mathrm{c}}\) and \(\tau = T_{\mathrm{c}} / T\). The values of the parameters \(n_i\), \(t_i\), \(d_i\), \(l_i\), \(\eta_i\), \(\beta_i\), \(\gamma_i\), and \(\epsilon_i\) are given in Table~2 of Ref.~\cite{Thol2016}.

\end{document}